\documentclass[floatfix,amsmath,amsfonts,twocolumn]{revtex4}  
\usepackage{enumitem}
\usepackage{amsmath}
\usepackage{graphicx}
\usepackage{epsfig}
\usepackage{bm}
\usepackage{bbold}
\usepackage{amssymb}
\usepackage{color}

\newcommand{\pt}{{\partial_t}}

\begin{document}

\title{Excitation of Wannier-Stark states in a chain of coupled optical resonators with linear gain and nonlinear losses}

\author{A. Verbitskiy}
\affiliation{School of Physics and Engineering, ITMO University, Kronverksky Pr. 49, bldg. A, St. Petersburg, 197101, Russia}

\author{A. Yulin}
\affiliation{School of Physics and Engineering, ITMO University, Kronverksky Pr. 49, bldg. A, St. Petersburg, 197101, Russia}

\date{\today}

\begin{abstract}

In this paper we theoretically study the nonlinear dynamics of Wannier-Stark states in the dissipative system consisting of interacting optical resonators, whose resonant frequencies depend linearly on their number. It is shown that the negative losses in some resonators can switch the system into a lasing regime with Wannier-Stark states serving as working modes. 
It is shown by extensive numerical simulations that there may be single-frequency stationary regimes as well as multi-frequency regimes. In the latter case Bloch oscillations can appear in the system.  
The possibility of selective excitation of Wannier-Stark states by the appropriate choice of the dissipation profile is investigated. A simple perturbation theory describing the quasi-linear regimes is developed and compared against the numerical results.


\end{abstract}

\maketitle

\section{Introduction}

Wannier-Stark ladders (WSL) continue to be of great interest to scientists in different fields of physics, such as, solid-state physics, condensed matter and quantum magnets \cite{interest_1, interest_2, interest_3}. The WSL effect consists in the presence of equidistant lines in the spectrum, which correspond to the eigenmodes of the system (Wannier-Stark states) \cite{discovery_1, discovery_2}. The beating between these states in time may result in periodic motion - Bloch oscillations (BOs) \cite{discovery_3, discovery_4, discovery_5}.

BOs was initially predicted in solid state physics. However, the experimental observation of this phenomenon in solid is a challenging problem. This explains why it took so many years to confirm the effect experimentally \cite{BO_exp}. However, it is worth noting here that BOs are a very common phenomena, so it has been found in a large variety of physical systems such as atomic systems \cite{atomic1, atomic2, atomic3, atomic4, atomic5, atomic6}, lasers \cite{laser}, coupled LC circuits \cite{LC}, mechanical systems \cite{mech1, mech2, mech3, mech4}, plasmonic \cite{plasmonic1, plasmonic2, plasmonic3, plasmonic4, plasmonic5, plasmonic6} or exciton-polariton systems \cite{exciton1, exciton2, exciton3}.

The advantage of the optical systems is that the experiments allowing to observe the aforementioned effects are less involving compared to the experiments in solid state physics. So the theoretical prediction of optical WSL and BOs \cite{prediction_1, prediction_2,BO_optics1, BO_optics2, BO_optics3, BO_optics4} has been quickly followed by the experimental demonstrations.

One of the first experimental study on observing optical WSL was done in \cite{WSL_exp1}. Here, using a chirped Moire grating, a system concept similar to the electronic one, was implemented. No less intriguing evidences of the existence of WSL are presented in the works \cite{WSL_exp2, WSL_exp3, WSL_exp4}. BOs have also been experimentally detected in the optical range \cite{BO_optics_exp1, BO_optics_exp2, BO_optics_exp3, BO_optics_exp4, BO_optics_exp5}. For the complete review of the research on BOs and related phenomena see \cite{BO_optics_review}. 

The presence of dissipation, pump and nonlinear effects in optical systems (for example in arrays of interacting nonlinear optical cavities) calls for the generalization of the theory of BOs for the case of nonlinear dissipative systems. Let us remark that such optical systems as arrays of microlasers can be seen as promising sources of coherent radiation \cite{microlasers_review_1, microlasers_review_2, microlasers_review_3, microlasers_review_4, microlasers_review_5, microlasers_review_6}. Thus the study of these systems are not only of fundamental but also of practical interest. 

In this paper we aim to study the regimes of radiation generation in one-dimensional systems of coupled optical cavities, where each of the resonators sustains only one mode, and its structure is determined by the material and geometry of the resonator. The sketch of the described system is shown in Fig. 1. To make our system to be of Bloch kind we introduce linear dependence of cavities frequency on their numbering index. The resonators become microlasers if they posses positive linear gain caused by the population inversion of their electronic excitations. This can normally be done either by optical or electric pump.   

However the interactions of different Wannier-Stark (WS) states can make the dynamics complicated resulting in the multi-frequency and, possibly, even chaotic behaviour. Below we consider in detail different regimes of WS lasers, switching from single frequency to multi-frequency regimes and the appearance of BOs. To explain the behaviour of the systems in the vicinity of the lasing threshold we develop a perturbation theory.  We consider this work as a proof of concept rather then a discussion of the optimal experimental system. Therefore we utilize the simplest model of the lasing cavities. We would like to mention that it can happen that for real experiment the scheme and, correspondingly, the theoretical model should be elaborated. 

\begin{figure}[ht]
 \begin{center} \includegraphics[width=0.45\textwidth]{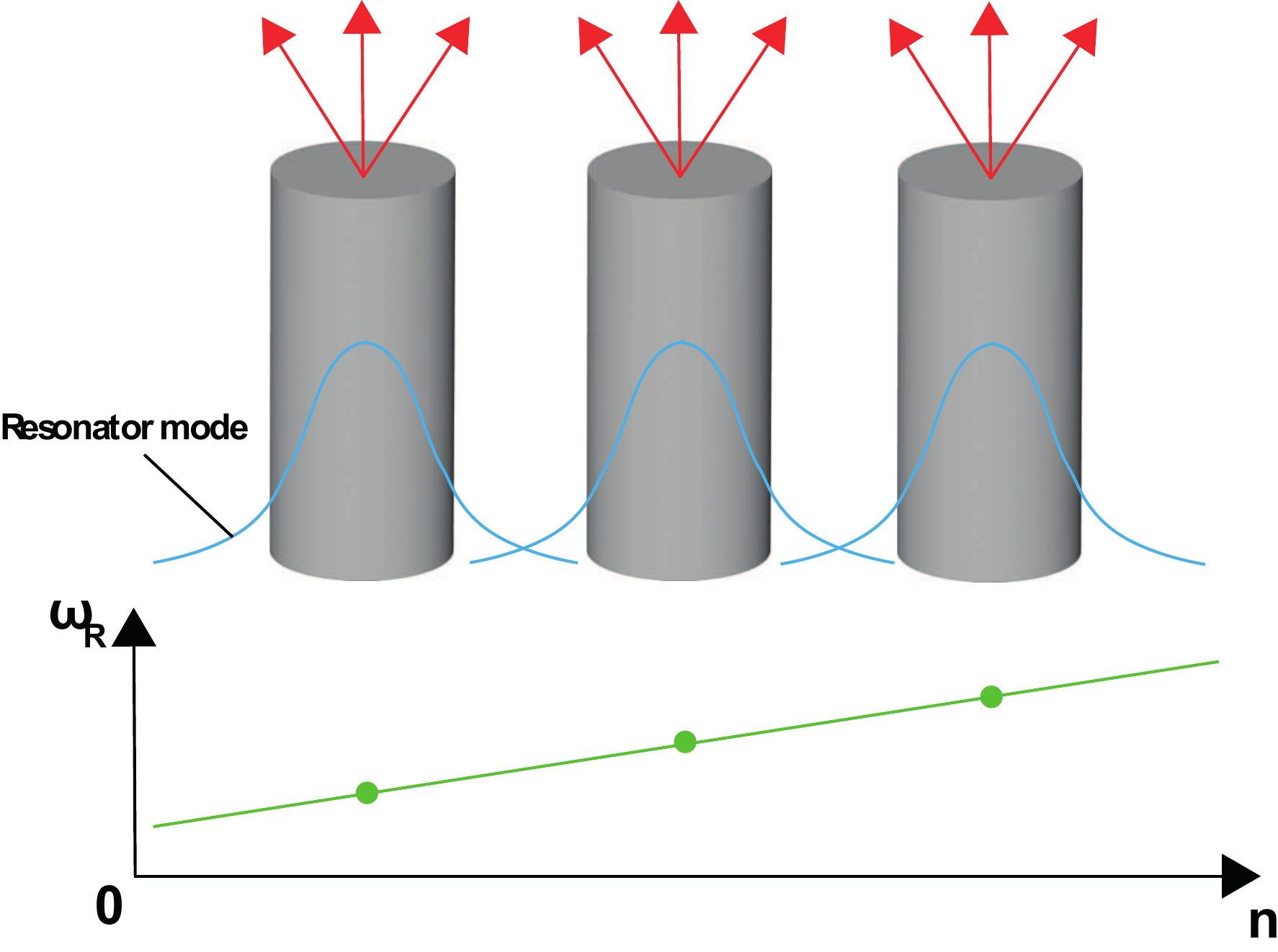}
  \caption{(Color online) The sketch of the considered system.}
  \end{center}
\end{figure}

To describe the dynamics of light in  microresonators we use a well known discrete model for the slow varying in time complex amplitudes $U_n(t)$ of the modes of the individual resonators \cite{Example_model1,Example_model2,Example_model3,Example_model4,Example_model5,Example_model6,Example_model7,Example_model8,Example_model9,Example_model10,Example_model11,Example_model12}. For sake of mathematical convenience we use dimensionless variables:
\begin{widetext}
\begin{eqnarray}
i \pt U_n + \sigma (U_{n+1}+U_{n-1}-2U_{n}) + \mu n U_n + i \gamma_n U_n + i\beta_n|U_n|^2 U_n = 0,
	\label{eq:main}
\end{eqnarray}
\end{widetext}
where $t$ is normalized time, $n$ is the index enumerating the resonator, $\sigma$ is the coupling strength between the resonators which can be set $\sigma=1$ without loss of generality, $\mu$ accounts for the dependence of the resonant frequency on the resonator index, $\gamma_n$ and $\beta_n$ are the linear and nonlinear losses correspondingly. Both $\gamma_n$ and $\beta_n$ can be different for different resonator. Let us remark, that here we consider a simple, but physically meaningful case, assuming that the nonlinear effects changes the effective losses but not the resonant frequencies of the individual resonators. We acknowledge that the effect of the nonlinear correction of the resonant frequencies can be of importance but it requires a special consideration and will be done elsewhere.

Strong enough incoherent pump can not only change the linear losses but make them negative. This means that this pump can make an individual cavity to be a laser. However considering the system of the resonators we have to calculate the effective gain of the supermodes of the system rather then the effective gain of the individual resonator.  Very roughly we can consider the stationary states as a balance between the effective gain and the effective nonlinear losses calculated for the WS state. Let us remark here, that there may be a case where the nonlinear losses are present only in the pumped cavities, and the case where the nonlinear losses are present in all resonators. Below we will show that in these two cases the dynamics of the WS modes is different.  

The paper is structured as follows. To do a systematic study of the problem we start with the simplest case where only one resonator is pumped. This is done in Section II of the paper. 
In Section III we show that simultaneous excitation of several resonators makes the dynamics richer giving rise to multi-frequency regimes including self-sustained BOs. In Section IV the mode selection is considered. It is shown that efficiency of mode excitation depends on pump profile and controlling the shape of the pump it is possible to extend range of intensities where the single-frequency regime takes place. The main findings of the work are briefly discussed in the Conclusion.

\section{Systems excited by linear gain in only one resonator}

We start our consideration with a simple case where $\gamma_n$ is negative in only one resonators with $n=0$, everywhere else $\gamma_n$ is a positive constant. This means that we have a linear amplification in the resonator $n=0$ and linear losses in all other resonators. 

We choose the linear losses in the form $\gamma_n=\gamma$ for $n \neq 0$ and $\gamma_0=\gamma - a$, where $a$ is pump amplitude, and study the dynamics of the system by numerical simulations.
The numerical simulations reveal that only trivial solution $U_n=0$ is possible until the linear gain $a$ is below a lasing threshold that depends of the parameters of the system $\gamma$ and $\mu$. If the threshold is exceeded then the growth of eigenmodes of the system is observed. If the dissipative and nonlinear terms are small then the growing modes can be very accurately approximated by the WS states known analytically  for the equation (\ref{eq:main}) in the conservative limit $\gamma_n \rightarrow 0$, \cite{BO_optics1}. The eigenvalues of the WS states form equidistant spectrum $\omega_m=\mu m$ with the eigenfunctions $W_{n-m}=J_{n-m}(\frac{2\sigma}{\mu})$,  index $m$ enumerate the eigenstates. We use the WS states normalized so that $\sum_n W_{n-m}^2=1$.

A simple perturbation theory can be developed if the dissipation is so small that it does not affect the spatial structure of the eigenstates. The quantity $E=\sum_n |U_n|^2$ (energy of the field in the system) is a conserving quantity if $\gamma_n=0$ and $\beta_n=0$. If $\gamma_n$ and $\beta_n$ are nonzero but small then the field in the system can be found in the form $U_n^{(m)}=A_m(t) W_{n-m} \exp(i\mu m t)$, where $A_m(t)$ is the time dependent complex amplitude of the $m$-th WS state. Substituting this into (\ref{eq:main}), multiplying by $W_{n-m}$ and calculating the sum over $n$ we obtain the ordinary differential equations for $A_m$:
\begin{eqnarray}
\pt A_m= - \Gamma_m A_m - {\cal B}_m |A_m|^2 A_m,
\label{eq:dyn_ampl_single_mode}
\end{eqnarray}
where $\Gamma_m=\sum_n \gamma_n W_{n-m}^2$ and ${\cal B}_m=\sum_n \beta_n W_{n-m}^4$ are the effective linear and the nonlinear losses for the $m$-th mode.

For pure dissipative nonlinearity (the nonlinearity affects only the effective losses but not the resonant frequency of the cavities) the equations (\ref{eq:dyn_ampl_single_mode}) can be re-formulated as a set of equations for the intensities $I_m = |A_m|^2 $:
\begin{eqnarray}
\pt I_m= 2 (- \Gamma_m I_m - {\cal B}_m I_m^2).
\label{eq:dyn_int_single_mode}
\end{eqnarray}

For our choice of $\gamma_n = 
\gamma - a \delta_{0n}$ ($\delta_{ij}$ is
Kronecker symbol) the sum in the expression for the effective linear losses $\Gamma_m$ can easily be calculated analytically: 
\begin{eqnarray}
\Gamma_m=\gamma - a W_{-m}^2.
	\label{eq:growth_rate}
\end{eqnarray}
The distributions of the intensity in the WS states are symmetric and have two main maxima situated symmetrically with respect to center of the mode. This fact means that if the system is excited by linear gain only in one resonator, then there are two fastest growing modes with the same increment. For the parameters we use for the numerical simulations the indexes of the fastest growing modes are $m_{max}=\pm 8$.

Now let us compare the results of the perturbation theory against the direct numerical simulations of the master equation (\ref{eq:main}). It is natural to introduce an effective linear gain of the mode as $-\Gamma_m $. The effective linear gains extracted from the numerical simulations and calculated by formula (\ref{eq:growth_rate}) are shown in Fig. 2(a) as functions of the pump amplitude $a$. It is seen that in the vicinity of the lasing threshold, where the dissipative terms can be considered as small corrections, the results of the perturbation theory fit the numerical simulations very well. 

Let us remark that the complex frequencies of the modes can be found by analysing the linearized equation for the amplitudes $U_n$:
 \begin{eqnarray}
i \pt U_n + \sigma (U_{n+1}+U_{n-1}-2U_{n}) + \mu n U_n + i \gamma_n U_n  = 0.
	\label{eq:main_lin}
\end{eqnarray}
Then looking for a solution in the form $U_n (t)= V_n \exp (i \omega t) $ we obtain an eigenvalue problem:  
 \begin{eqnarray}
\omega V_n = \sigma (V_{n+1}+V_{n-1}-2V_{n}) + \mu n V_n + i \gamma_n V_n.
	\label{eq:main_eigval}
\end{eqnarray}
The real part of $\omega$ is the frequency of the eigenmode, the imaginary part - its dissipation rate, the eigenvector $V_n$ describes the structure of the eigenmode. In the absence of the dissipative term the eigenstates are the conservative WS states discussed above. The solution of the spectral problem allows to find the exact solutions for the eigenstates in the dissipative case. We solved the spectral problem numerically to confirm that the dissipative terms does not affect much the structure of the eigenmodes.

It is also instructive to compare effective linear gains of different WS states. The numerically found $-\Gamma_m$ for six fastest growing modes are shown in Fig. 2(b) as function of the pump amplitude $a$. One can see that the fastest growing modes with the lowest lasing threshold, for our choice of parameters, are the modes with $m=\pm 8$, the second and the third fastest growing modes have the indexes $m=\pm 9$ and $m=\pm 2$ correspondingly. 

\begin{figure}[ht]
 \begin{center} \includegraphics[width=0.45\textwidth]{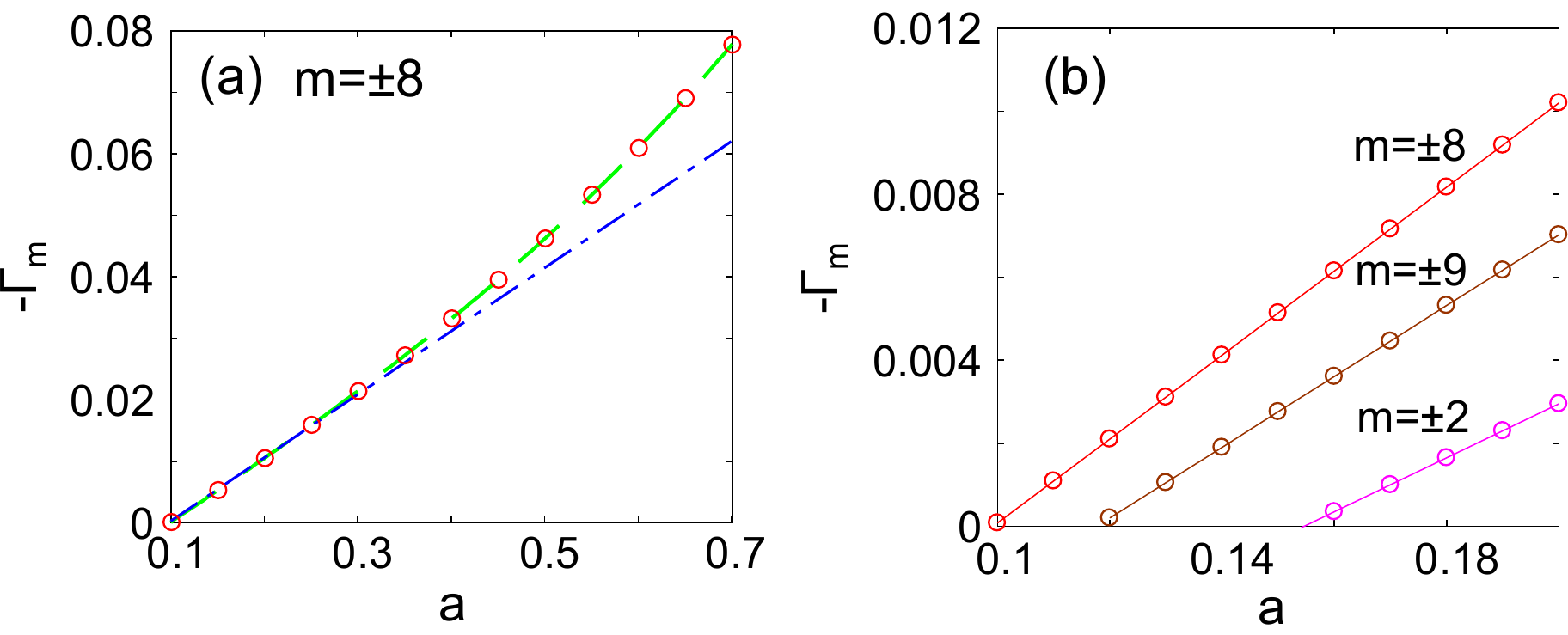}
  \caption{(Color online) (a) The dependencies of the effective linear gains $-\Gamma_m$ of the fastest growing Wannier-Stark states with $m=\pm 8$ on the pump amplitude $a$ obtained by numerical simulation (red circles), via the perturbation method (dash-dotted blue line) and via the eigenvalues (dashed green line). (b) The dependence of the effective linear gains $-\Gamma_m$ of Wannier-Stark states with $m=\pm 8$ (red circles), $m=\pm 9$ (brown circles) and $m=\pm 2$ (magenta circles) on the pump amplitude $a$ obtained by numerical simulation. The solid lines are guides for eyes. Parameters are: $\mu=0.2$, $\gamma=0.01$. 
  }
  \end{center}
\end{figure}

The intensity of the stationary states $\overline{I_m}$ forming in the system can easily be found from (\ref{eq:dyn_int_single_mode}):
\begin{eqnarray}
\overline{I_m}=\frac{-\Gamma_m}{{\cal B}_m}.
	\label{eq:st_int}
\end{eqnarray}
The dependencies on the pump $a$ of the stationary intensities of three pairs of WS states having the highest effective linear gains are shown in Fig. 3 for two cases: (a) for the case where the nonlinear losses are nonzero only in the excited resonator with $n=0$: $\beta_0=\beta$ and (b) for the case of spatially uniform nonlinear losses: $\beta_n=\beta$. It is interesting to note that the stationary intensities can be higher for modes with lower effective linear gains, see Fig. 3(a). This can be explained by the fact that in the case where the nonlinear losses are nonzero only in the excited resonator the modes with the highest effective linear gains have the highest nonlinear losses so that their ratio (\ref{eq:st_int}) is less compared to that of the modes with the lower effective linear gains. 

We performed numerical simulations that reveal that only one pair of the WS states with the highest effective linear gain is dynamically stable for small pump intensities. The dependencies of the stationary intensities of the WS states extracted from the numerical simulations are shown in Fig. 3. The comparison of the analytics and the numerics shows that the agreement between the perturbation theory and the numerical simulations is good for the low intensities of the pump.

\begin{figure}[ht]
 \begin{center} \includegraphics[width=0.45\textwidth]{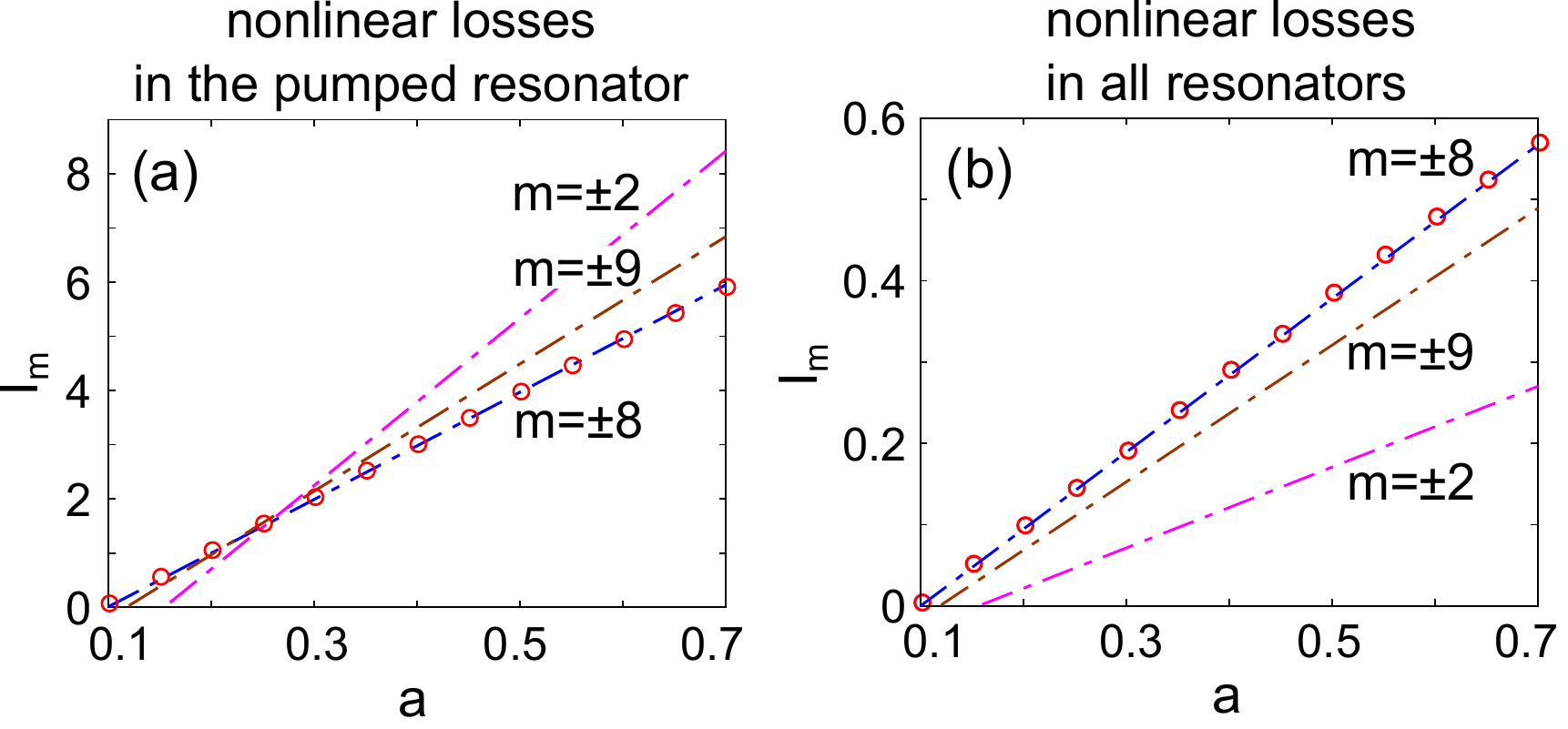}
  \caption{(Color online)   
  The dependencies of the stationary intensities of different Wannier-Stark states $I_m$ on the pump amplitude $a$ obtained via the perturbation method (dash-dotted lines) and by numerical simulation (circles) (a) for the case where the nonlinear losses are nonzero only in the excited resonator with $n=0$: $\beta_0=\beta$ and (b) for the case of spatially uniform nonlinear losses: $\beta_n=\beta$. Parameters are: $\mu=0.2$, $\gamma=0.01$, $\beta=1$. 
  }
  \end{center}
\end{figure}

It is instructive to study the dynamics when the initial conditions are taken in the form of noise of small intensity. As it is mentioned above the modes start growing when the pump exceeds some threshold. We choose the pump exceeding only the threshold for the fastest growing modes. So, for our choice of the parameters, only the modes $m=\pm 8$ grow. The numerical simulations show that in the case of the nonlinear losses present only in the pumped resonator we see the formation of a single frequency stationary state in the form of WS state with $m=8$ or $m=-8$. The probability of the formation of each of the states is $\frac{1}{2}$. The formation of the stationary states is illustrated in Fig. 4(a), (b), (d) and (e).

\begin{figure*}[ht]
 \begin{center}  \includegraphics[width=0.9\textwidth]{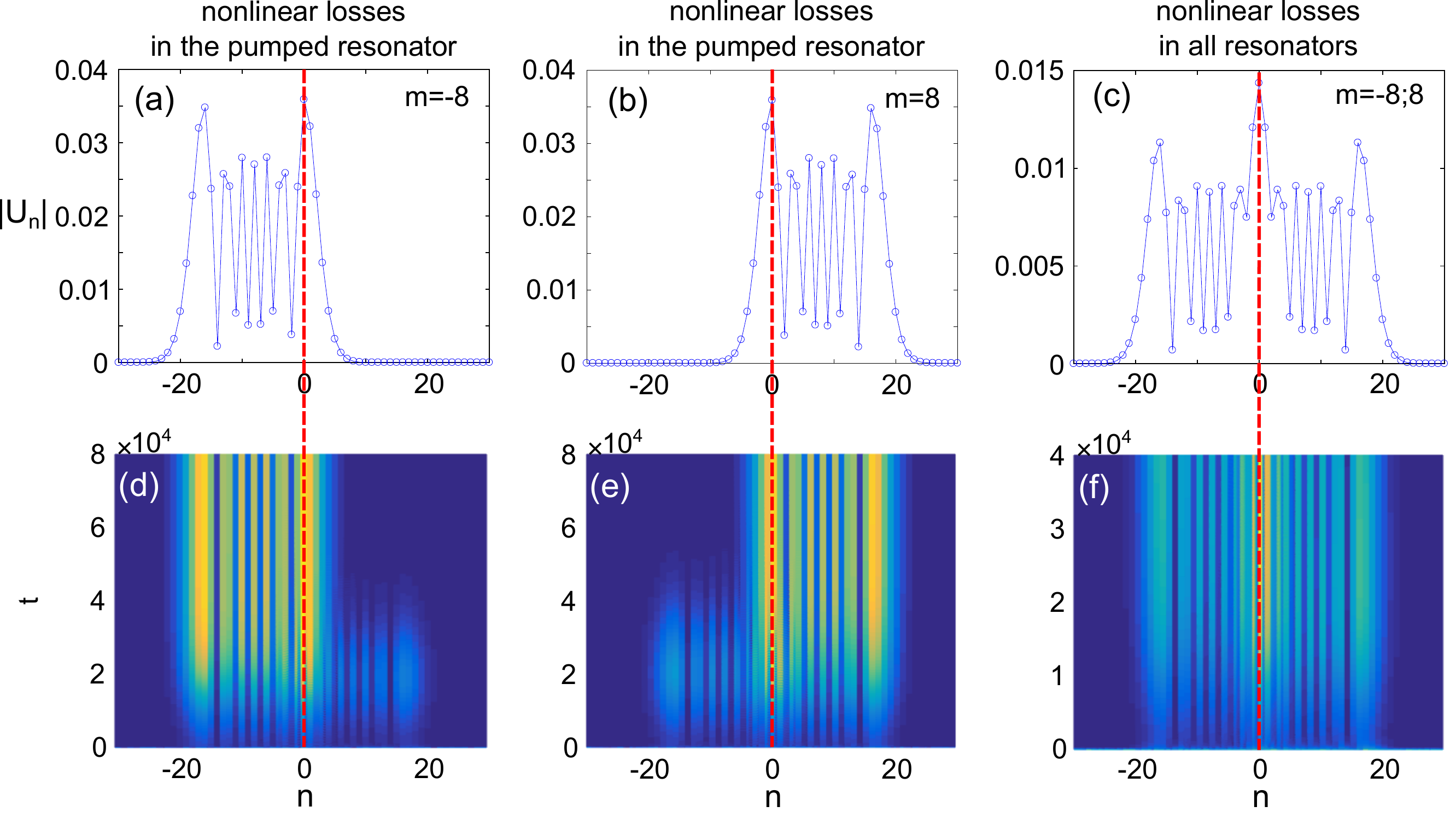}
  \caption{(Color online) The stationary states $|U_n|$  in the form of WS states with (a) $m=-8$, (b) $m=8$, (c) $m=-8$ and $m=8$ (time average field value); and their evolutions of the field module $|U_n(t)|$ (d), (e) and (f) correspondingly. The pump amplitude slightly exceeds the excitation threshold. The nonlinear losses are nonzero only in the excited resonator with $n=0$: $\beta_0=\beta$ for (a), (b), (d) and (e); the spatially uniform nonlinear losses: $\beta_n=\beta$ for (c), (f). The blue circles correspond to the resonators, the solid blue lines are guides for eyes, the dashed red lines correspond to the pumped resonator with $n=0$. Parameters are: $\mu=0.2$, $\gamma=0.01$, $a=0.1$, $\beta=1$.
  }
  \end{center}
\end{figure*}

Different regimes of the stationary states formation occur if the nonlinear losses are distributed evenly in the system. The excitation thresholds, of course, remain the same, but the stationary state forming from a weak noise varies periodically in time. Very close to the excitation threshold the stationary state can be seen as a superposition of the WS states with $m=8$ and $m=-8$, consequently the stationary state contains the temporal harmonics with the frequencies equal to the eigenfrequencies of the WS states. The formation of such a state is illustrated in Fig. 4(c),(f).

To explain such system behavior we expanded the perturbation theory described above by writing the equations for the amplitudes $A_{\pm}$ 
of two interacting modes $m=\pm \tilde m$ having the highest effective linear gains. So we sought the field in the form $U_n=A_{+}W_{n-\tilde m}\exp(i \tilde m t) +A_{-}W_{n+\tilde m}\exp(-i \tilde m t)$. Substituting this ansatz into (\ref{eq:main}) and projecting on the eigenstates we obtain the equations for $A_{\pm}$. 
These equations can be reduced to the equations for the intensities $I_{\pm}$ in the same way as was done for (\ref{eq:dyn_int_single_mode}):
\begin{eqnarray}
\pt I_{+}= -2 (\Gamma + {\cal B} I_{+}+\tilde {\cal B} I_{-})I_{+},
	\label{eq:dyn_int_two_modes1} \\
\pt I_{-}= -2 (\Gamma + {\cal B} I_{-}+\tilde {\cal B} I_{+})I_{-},
	\label{eq:dyn_int_two_modes2} 
\end{eqnarray}
where $\tilde {\cal B}=2\sum_n \beta_n W_{n-\tilde m}^2W_{n+\tilde m}^2$, $\Gamma=\Gamma_{\pm \tilde m}$. Deriving these equations we assumed that the difference of the eigenfrequencies of the states is large and so we can safely disregard the frequently oscillating terms.

Let us analyse the fixed points of the dynamical system (\ref{eq:dyn_int_two_modes1})-(\ref{eq:dyn_int_two_modes2}). For $\Gamma>0$ there is only a trivial solution $I_{\pm}=0$. 
For the negative losses (and, correspondingly positive gain) there are four solutions: $I_{\pm}=0$; $I_{+}=0$, $I_{-}=\frac{-\Gamma}{{\cal B}}$; $I_{-}=0$, $I_{+}=\frac{-\Gamma}{{\cal B}}$ and $I_{\pm}=\frac{-\Gamma}{{\cal B}+\tilde {\cal B}}$. 

It is straightforward to study the stability of the states writing the linearized equations for small perturbations $\xi_{\pm}$ of the intensities $I_{\pm}$ and finding the eigenvalues governing the evolution of the perturbations. The trivial state is, of course, always unstable $\lambda_{1,2}=-2\Gamma$. The second and the third states have the eigenvalues $\lambda_1=-2\Gamma(1-\frac{\tilde {\cal B}}{{\cal B}})$ and $\lambda_2=2\Gamma$. $\lambda_2$ is always negative for $\Gamma<0$, $\lambda_1$ is negative for ${\cal B}<\tilde {\cal B}$ and positive otherwise. This means that this state can be either a stable node for ${\cal B}<\tilde {\cal B}$ or a saddle for ${\cal B}>\tilde {\cal B}$. The last state $I_{\pm}=\frac{-\Gamma}{{\cal B}+\tilde {\cal B}}$ has the eigenvalues $\lambda=\frac{2\Gamma}{{\cal B}+\tilde {\cal B}}  \left( {\cal B}\pm\tilde {\cal B}  \right)$. From this we can conclude that this state is stable (a stable node) for ${\cal B}>\tilde {\cal B}$ or unstable (saddle) for ${\cal B}<\tilde {\cal B}$.

So the stability analysis tells us that if ${\cal B}>\tilde {\cal B}$ there is only one stable stationary state $I_{\pm}=\frac{-\Gamma}{{\cal B}+\tilde {\cal B}}$ for the system (\ref{eq:dyn_int_two_modes1})-(\ref{eq:dyn_int_two_modes2}). For ${\cal B}<\tilde {\cal B}$ there are two stable states $I_{+}=0$, $I_{-}=\frac{-\Gamma}{{\cal B}}$ and $I_{+}=\frac{-\Gamma}{{\cal B}}$, $I_{-}=0$.

Now let us estimate the values $\tilde {\cal B}$ and $ {\cal B}$. For the case, when only $\beta_0 \neq 0$, $ {\cal B}=\sum_n \beta_n W_{n-\tilde m}^4=W_{\tilde m}^4$ and $\tilde {\cal B}=2\sum_n \beta_n W_{n-\tilde m}^2W_{n+\tilde m}^2=2W_{\tilde m}^4$. This means that $\tilde {\cal B}=2{\cal B}$ and, as our stability analysis shows, in this case the stable stationary states are $I_{+}=0$, $I_{-}=\frac{-\Gamma}{{\cal B}}$ and $I_{+}=\frac{-\Gamma}{{\cal B}}$, $I_{-}=0$. Only a stable state can be observed as a stationary state in numerical simulations and this explains why for this choice of $\beta_n$ we see the formation of either one or another WS state. 

In the case when $\beta_n=\beta$ the ration of  ${\cal B}$ and $\tilde {\cal B}$ can be different. The coefficient $\tilde {\cal B}$ depends on the overlap of the intensity distributions of the states $W_{n\pm \tilde m}$ and it is easy to see that this overlap decreases with the increase of the width of the WS state defined as $H=\sqrt{\sum_n W_{n}^2 (n-n_c)^2}$, where $n_c$ is the center of the WS state. Fig. 5 shows the dependencies of ${\cal B}$ and $\tilde {\cal B}$ on the width of the states $H$, (the width of the state $H$ is determined by $\mu$).

\begin{figure}[ht]
 \begin{center} \includegraphics[width=0.45\textwidth]{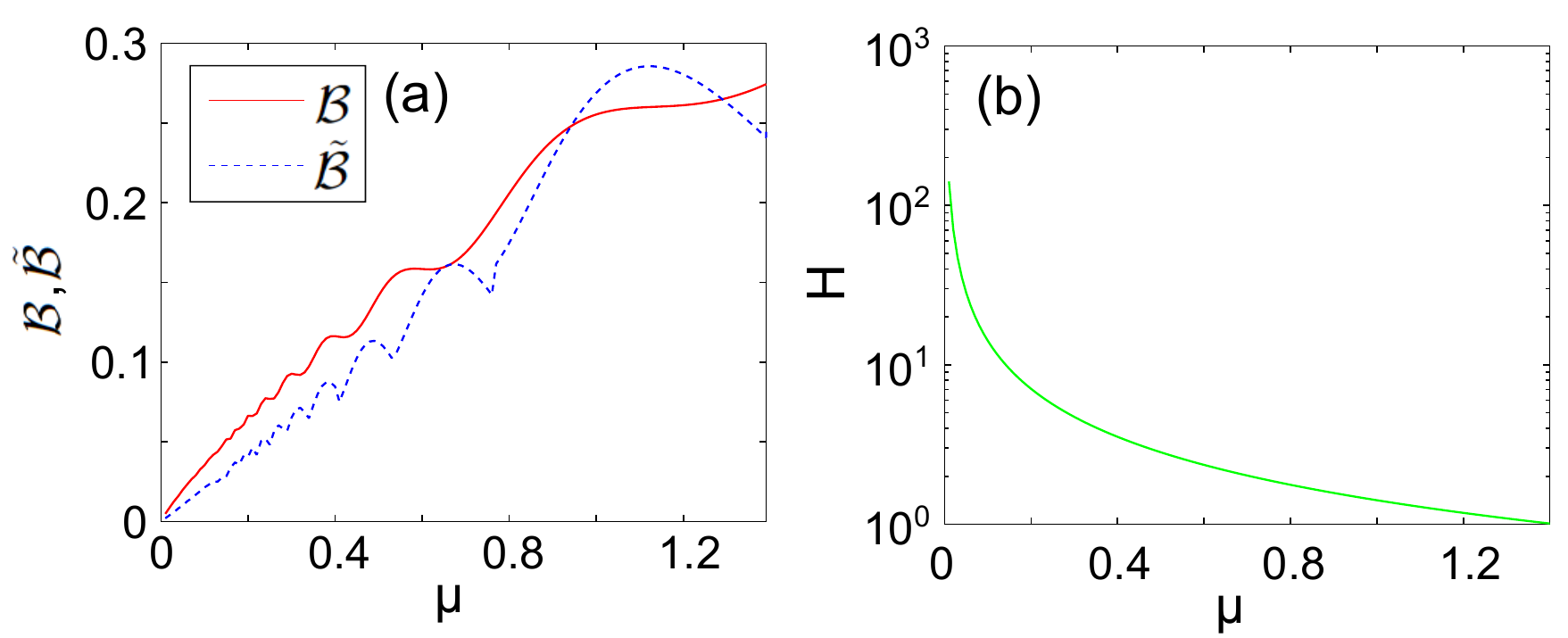}
  \caption{(Color online) (a) The dependencies of ${\cal B}$ and $\tilde {\cal B}$ on $\mu$ for the spatially uniform nonlinear losses: $\beta_n=\beta=1$. (b) The dependence of the width of the WS state $H$ on $\mu$.}
  \end{center}
\end{figure}

For $\mu = 0.2$ used in our direct modelling the coefficients are ${\cal B}=0.07$ and $\tilde {\cal B}=0.04$. This means that in this case there is only one stable stationary state $I_{\pm}=\frac{-\Gamma}{{\cal B}+\tilde {\cal B}}$.  
So one can anticipate that in this case the final state consists of two WS states of the same intensity and oscillating with different frequencies. This fits the results of our numerical simulations perfectly, see Fig. 4(c). It is also worth noting that for $\mu>0.6$ the bands of $\mu$ where $\tilde {\cal B}>{\cal B}$ appear, see Fig. 5(a). This means that in these bands there should be two stable states $I_{+}=0$, $I_{-}=\frac{-\Gamma}{{\cal B}}$ and $I_{+}=\frac{-\Gamma}{{\cal B}}$, $I_{-}=0$ instead of the previously observed single state. This fact is confirmed by numerical calculations. 

We would like to note that the developed perturbation theory gives not only qualitative explanation of the observed effect, but allows to determinate the intensities of the two-components states with good precision. The dependencies of the intensity $I_{\pm}(t)$ extracted from numerical simulations are overlapped with that calculated by formulas (\ref{eq:dyn_int_two_modes1})-(\ref{eq:dyn_int_two_modes2}). It is seen that for low values of the linear gain $a$ the agreement is good.

\section{Lasing with the linear gain in several resonators}

To increase the radiation power it makes sense to create linear gain in several resonators. Let us first consider the case where the nonlinear losses are present only in the pumped resonators. 
At first we note that if the gain is evenly distributed in the pumped resonators then, as one can anticipate, the lasing starts at lower pump amplitudes for larger number of pumped resonators, see Fig. 6(a), showing the dependence of the total energy $E$ of the single-mode stationary state as a function of the pump amplitude $a$ for different number of pumped neighbouring resonators $M$. It can be seen from this figure that the stationary energy values $E$, obtained via the perturbation method (solid line) and by numerical simulations (circles) are in good agreement with each other for different $M$.

\begin{figure*}[ht]
 \begin{center}
\includegraphics[width=0.7\textwidth]{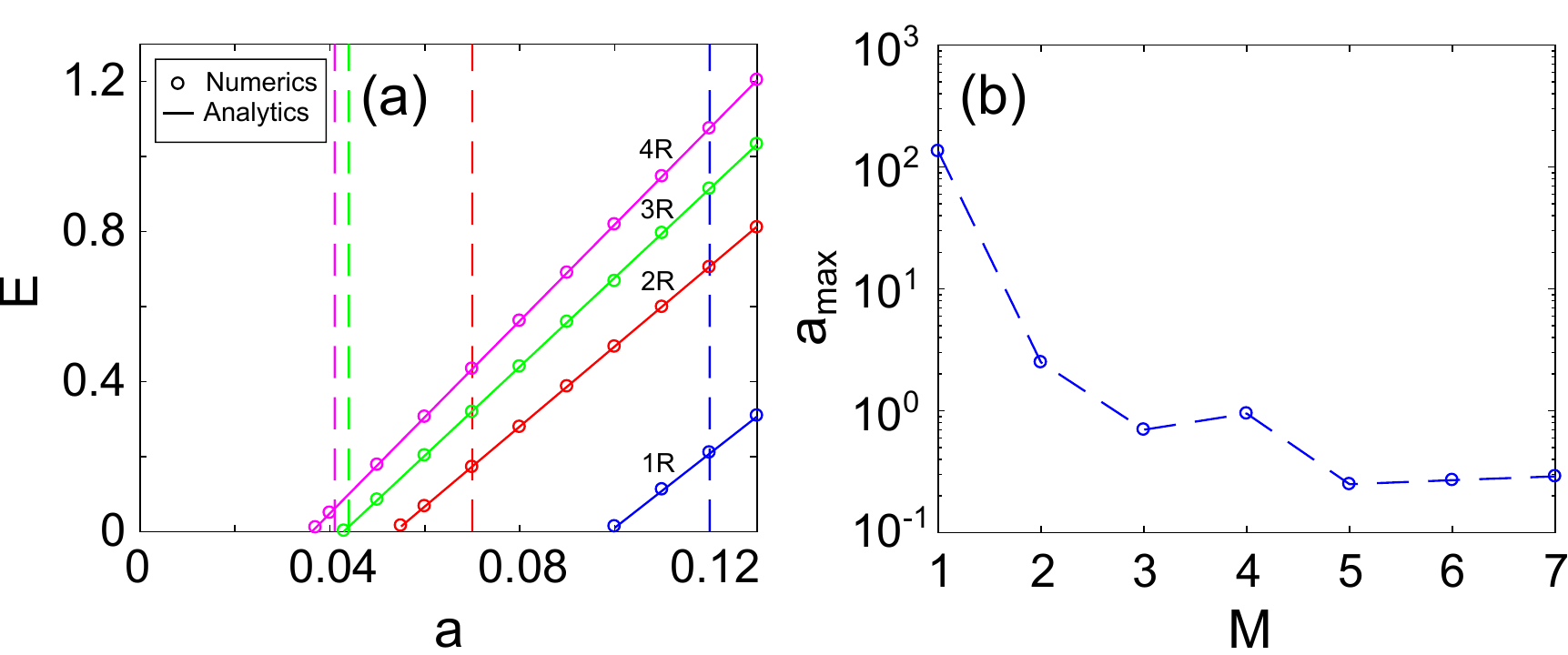}
  \caption{(Color online)   
  (a) The dependence of the total energy $E$ of the single-mode stationary state on the pump amplitude $a$ obtained via the perturbation method (solid line) and by numerical simulation (circles) for different number of pumped adjacent resonators $M$: 1R, 2R, 3R and 4R. Dashed lines show the threshold pump amplitudes at which next pair of Wannier-Stark states is excited. (b) The dependence of the maximum pump amplitude $a_{max}$ supporting a single frequency lasing regime on the number of pumped resonators $M$. The circles correspond to the analytical data, the dashed line is guide for eyes. 
  The nonlinear losses are nonzero only in the pumped resonators $\beta_M=\beta=1$.
  Parameters are: $\mu=0.2$, $\gamma=0.01$.} 
  \end{center}
\end{figure*}

The single frequency state is the only one possible solution within the range of the pumps $a_{th1}<a<a_{th2}$, where $a_{th1}$ is the excitation threshold of the fastest growing pair of WS modes and $a_{th2}$ is the excitation threshold of the second fastest growing pair. The simulations show that with an increase of the pumped resonators number $M$, the range of the existence of a single frequency solution decreases, see Fig. 6(a).

It is obvious that, at sufficiently large pump amplitudes $a$, the single-frequency stationary state becomes unstable and collapses. It is important to study how the maximum amplitude of the pump $a_{max}$ providing the single-frequency regime depends on the number of pumped sites $M$. This  dependence is shown in Fig. 6(b). It can be seen that the critical pump amplitude decreases with an increase of $M$. This behaviour can be explained by the fact that for wide pumps the overlap integrals defining the effective gain of the mode depend on the indexes of the modes weakly. This means that the modes have very similar excitation thresholds. Thus the single frequency regime exists only in a tiny range of the pumps between the excitation thresholds of the fastest and the second fastest growing modes.   

Let us now consider in more detail the dynamics of the system with the pump in three neighbouring resonators.  The numerical simulations show that in the vicinity of the threshold ($a \approx 0.043$) the radiation is monochromatic, see Fig. 6(a). The stationary states similar to those shown in Fig. 4(a) and (b) form with equal probabilities.

When the pump amplitude increase the threshold value for the next pair of WS states ($a \approx 0.044$) a multi-frequency regime appears in the system. For the pump slightly above this threshold the stationary state can be seen as a superposition of two WS states having different frequencies. Let us note that due to nonlinearity the temporal spectrum of the stationary state contains the whole set of combination frequencies but in the case of the weak nonlinearity there are two dominating frequencies corresponding to the eigenfrequencies of the modes. So this case is very similar to the case where the gain is present in only one resonator but the nonlinear losses are present in all resonators of the system, see Fig. 4(c). 

At the higher levels of the pump the multimode regime changes. The numerical simulations reveal that a complex states resembling BOs appear, see Fig. 7(a), (b). Their temporal spectra are shown in the panels (e), (f) of the same figure. The spectra contain four harmonics for the pump $a=0.06$ corresponding to WS states with the indexes $m = \pm 8$ and $m = \pm 9$, see Fig. 7(e). For the higher pump more temporal harmonics appear, see Fig. 7(f) showing the spectrum for $a=0.2$.

It is interesting to note that for the pumps exceeding a threshold level ($a \approx 0.13$ for our parameters) the intensity distribution becomes asymmetric, compare Fig. 7(a) and Fig. 7(b). After the symmetry breaking one pair of frequencies survives without serious changes but the other transforms into several spectral lines signalling the excitation of many WS states. So the right snaking patterns in panel (f) contains more harmonics and thus the BOs produced by the eigenstates belonging to this part of the spectrum become smoother.

\begin{figure*}[ht]
 \begin{center}
\includegraphics[width=0.95\textwidth]{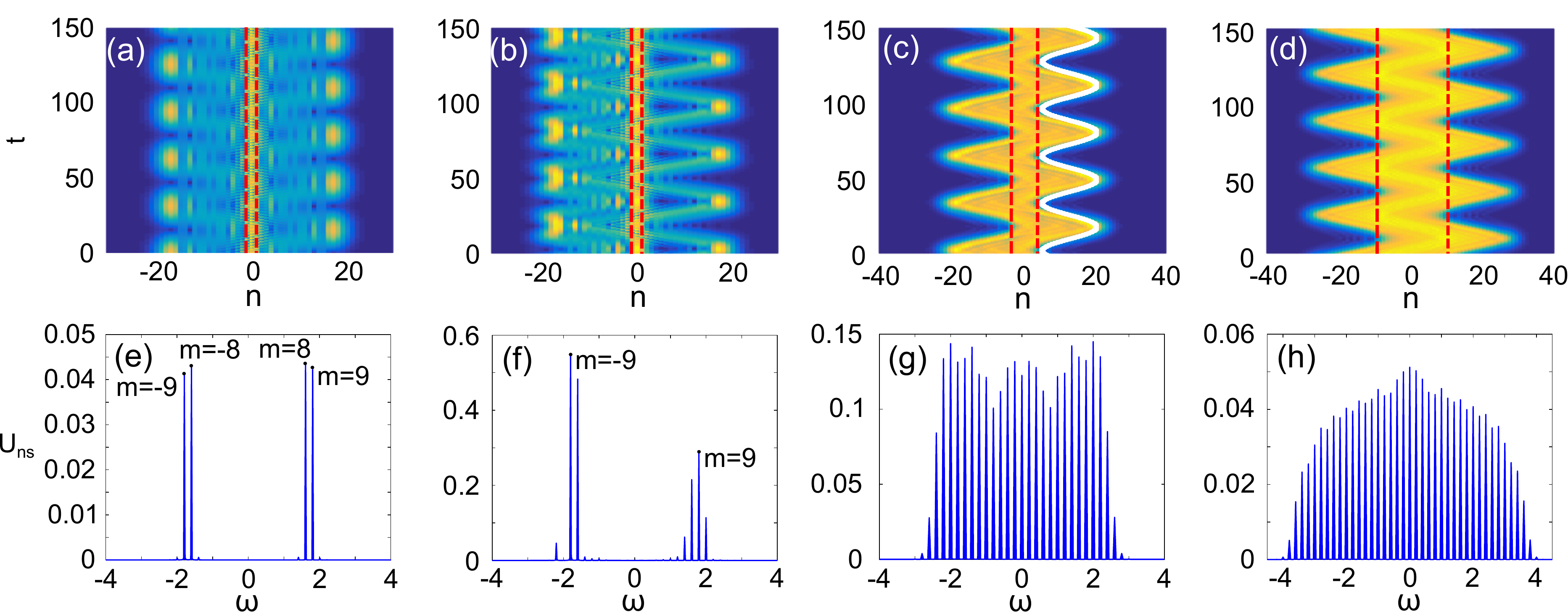}
  \caption{(Color online) The evolution of the field module $|U_n(t)|$ and its spectrum, defined as $U_{n \, s}(\omega)=\sum_n| \int_{-\infty}^{\infty} U_n(t)\cdot e^{-i\omega t}dt|^2$, for the multi-frequency stationary state: (a), (e) for $M=3$, $a=0.06$; (b), (f) for $M=3$, $a=0.2$; (c), (g) for $M=7$, $a=0.17$ and (d), (h) for $M=21$, $a=0.06$. The dashed red lines show the position of the pump area. In panel (c) the solid white line corresponds to the Bloch oscillations in the linear conservative system. The nonlinear losses are nonzero only in the pumped resonators $\beta_M=\beta=1$. Parameters are: $\mu=0.2$, $\gamma=0.01$.}
  \end{center}
\end{figure*}

For the higher numbers of the excited resonators the BOs  become smoother and have wide temporal spectrum, see Fig. 7(c), (d), (g), (h) showing the evolutions of the field amplitude and the temporal spectra for $M=7$ and $M=21$ excited resonators. To prove that the snaking pattern seen in panels (a)-(d) are related to the BOs we calculated analytically the trajectory of BOs for the linear conservative system and overimposed this curve with the panel (c). It is seen that the amplitude and the period of the BOs in the conservative counterpart of the considered system are very much similar to that observed in the direct numerical simulations. 

Let us remark here that we verified that the multi-frequency regimes are qualitatively the same regardless whether the nonlinear losses are present in all resonators or only in the excited ones. That is why in this paper we do not discuss the case of evenly distributed nonlinear losses.

\section{Mode selection}

For practical purposes it can be useful to increase the range of the pumps supporting the single-frequency regime. This problem seems especially important when the linear gain is created in many resonators which allows to increase the output power of the working mode.  To stabilize the single-frequency regime we suggest to profile the pump. The profiling allows to control the effective gain seen by the different modes and thus it is possible to achieve that one of the modes has the increment significantly higher than the increments of the other modes. 

We start with the case where only one of the resonators has positive linear gain and the nonlinear losses are present in this resonator. As it is discussed above, in this case there may form two different WS states with equal probability. Let us show that choosing the appropriate pump it is possible to provide that stationary states are predefined by the shape of the gain and thus do not depend on the initial conditions. For this we modify the effective losses in the individual resonators by adding some additional losses in the appropriate resonator, see Fig. 8(a, c), showing the distribution of the effective losses.  

Without the added losses there are two eigenmodes with the same effective linear gain but different frequencies and field distributions. Then we increase the losses in the resonator where one of the modes has intensity maximum but the intensity of the second mode is small in this resonator. This means that the added dissipation suppresses the effective linear gain of the first mode but does not change much the growth rate of the second mode. This way it is possible to achieve a controllable excitation of the desirable WS state. The evolution of the mode growing from weak noise is shown in Fig. 8 (a, c).

In Sec. II it is shown that in the case of the nonlinear losses present in all resonators the stationary state is a combination of two fastest growing WS states having different frequencies. By adding additional losses in one of the resonators it is possible to suppress one of the WS states. So in this case the modification of the pump profile can proved a single-frequency lasing. This is illustrated by Fig. 8(b, d), showing the evolution of the field with weak noise taken as the initial conditions. For $t<4 \cdot 10^4$ there is no additional losses and one can see the formation of a stationary state in the form superposition of two WS states. At $t=4\cdot 10^4$ we switch on the additional losses in the resonator $n=16$. This suppresses one of the WS state immediately and one can observe a stable single-frequency WS state.

\begin{figure*}[ht]
 \begin{center} \includegraphics[width=0.6\textwidth]{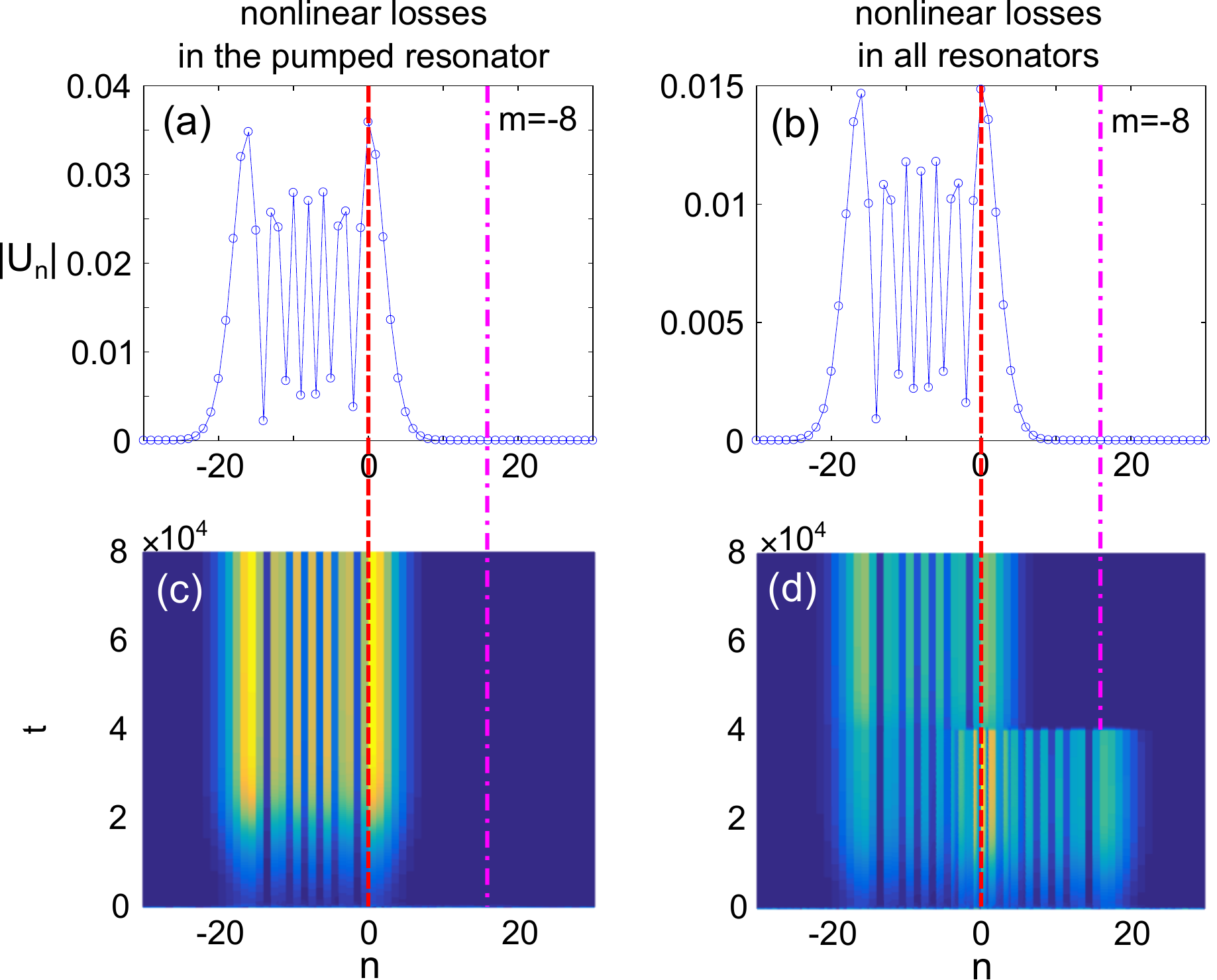}
  \caption{(Color online) The evolutions of the field amplitude $|U_n(t)|$ and stationary distribution of the amplitudes are shown for the cases where negative linear losses $\gamma_0=\gamma-a$ are only in the resonator with $n=0$  and additional losses $\tilde \gamma$ are added to the resonator with $n=16$, so that $\gamma_{16}=\gamma+\tilde \gamma$. The nonlinear losses are nonzero only in the excited resonator with $n=0$: $\beta_0=\beta$ for (a), (c); the spatially uniform nonlinear losses: $\beta_n=\beta$ for (b), (d). Panels (c,d) show the evolutions from a weak noise, panels (a,b) - the final distribution of the amplitudes. The blue circles correspond to the resonators, the solid blue lines are guides for eyes. The red dashed line marks the position of the resonator with the gain, the dash-dotted magenta line shows the position of the additional losses added to provide the selection of the mode. In the case of the nonlinear losses present in all resonator there is no additional losses at $t<4 \cdot 10^4$. So the formation of the hybrid state consisting of two WS states is seen in panel (d). The additional losses are switched on at $t=4 \cdot 10^4$ in the resonator $n=16$. One can see that this suppresses one of the WS states and the stationary state is a single frequency one with the amplitude profile corresponding to the fastest growing WS state, see panel (b). The parameters are: $\mu=0.2$, $\gamma=0.01$, $\tilde \gamma=0.03$, $a=0.1$, $\beta=1$.
  }
  \end{center}
\end{figure*}

To increase the power of the lasing mode it is natural to increase the area and the intensity of the pump. However, as it is discussed above, this makes single-mode regime difficult to observe. But, the question of single-frequency generation is most acute when the nonlinear losses are located in the excited resonators. This issue can be overcome by the profiling of the gain placing the pump in the resonators where the working mode has intensity maxima and adding the losses in the resonators where the other modes have big intensity. In numerical simulations we consider the case of $7$ pumped oscillators and different distribution of the effective losses, see Fig. 9(a-c).

\begin{figure*}[ht]
 \begin{center}
\includegraphics[width=0.95\textwidth]{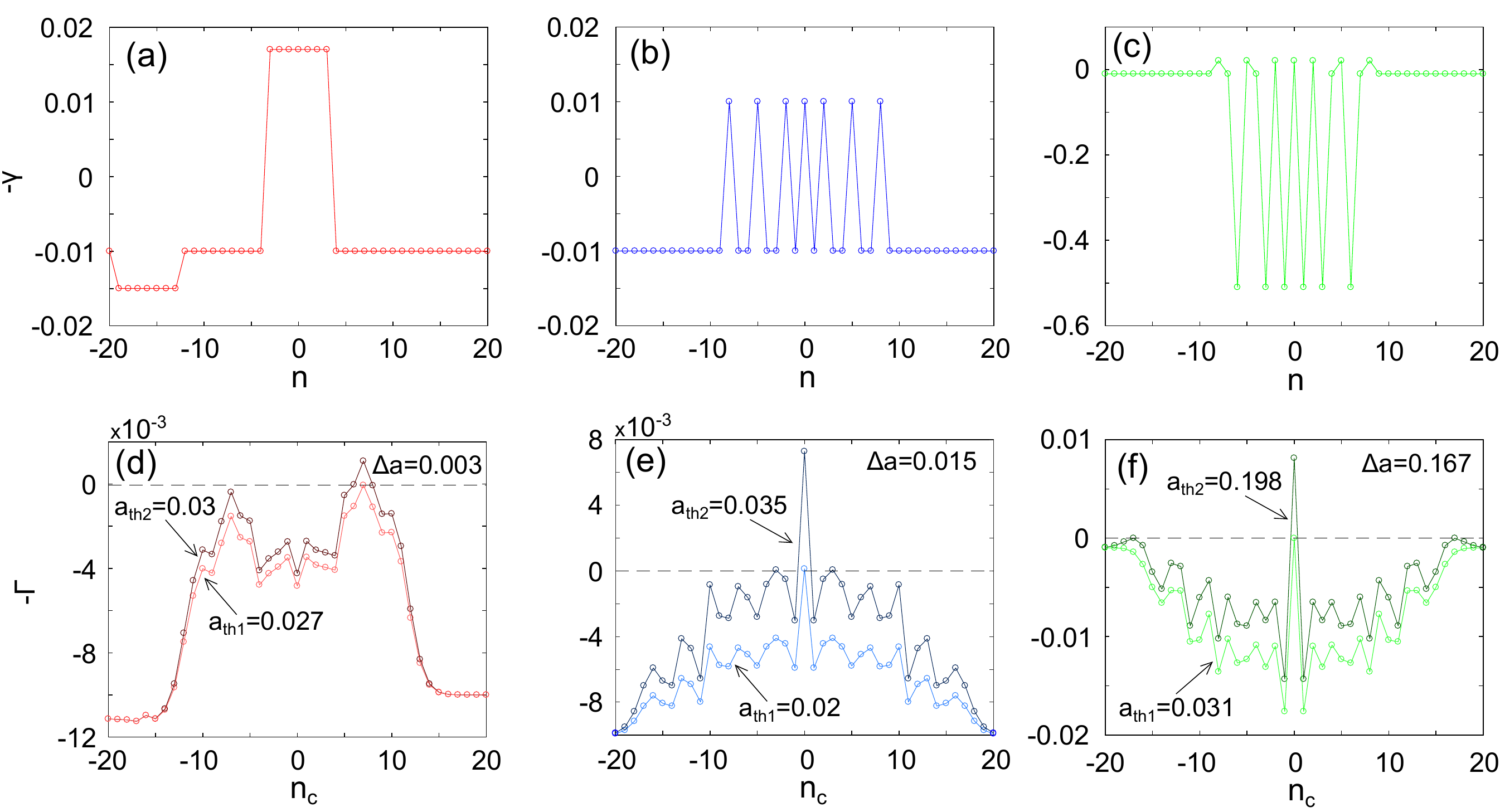}
  \caption{(Color online) The gain profiles $-\gamma_n$ with the pump amplitude, which is just above the lasing threshold: $a=0.027$ for (a), $a=0.02$ for (b), $a=0.031$ for (c). The additional losses for panel (a) $\tilde \gamma = 0.005$ and for panel (c) $\tilde \gamma = 0.5$. The circles correspond to the resonators, the solid lines are guides for eyes. (d)-(f) The dependences of the effective linear gain $-\Gamma$ on the Wannier-Stark state center position $n_c$ for two first threshold pump amplitudes $a_{th1,2}$ for the gain distributions (a)-(c) correspondingly. The circles correspond to the data from the perturbation method, the solid lines are guides for eyes. Parameters are: $\mu=0.2$, $\gamma=0.01$.
  }
  \end{center}
\end{figure*}

The effective gain created by these pumps depends on the index of the WS state. These dependencies are shown in panels (d-f) of Fig. 9. It is seen there that the effective gain for one of the modes is much greater than the gain for the other modes, which is demonstrated in panels (e, f). Moreover, it is seen that in wide range of the pump amplitudes the only one mode has a positive increment. So it can be anticipated that the fastest growing mode will define the stationary state.  

We checked by numerical simulations the hypothesis that if the only one mode has a positive increment then the final stationary state is a single frequency one. The numerical simulations fully confirm this prediction. At the same time, as expected, for the pump profile, shown in Fig. 9(a), the single-frequency generation range is extremely small ($E_{max}=0.04$). 

However, in numerics, we see that due to gain profiling single frequency lasing regimes continue to exist at sufficiently high pumps, where there are more than one growing modes. So if the pump is located in the resonators where working mode has intensity maxima, see Fig. 9(b), the existence range of the single-frequency regime increases by two orders of magnitude ($E_{max}\approx 4$). This behavior can be explained by the fact that this pump profiling provides a better selection of the working mode in terms of the effective gain difference, see Fig. 9(e).

It is possible to increase the single-frequency range even further by adding the additional losses in the resonators where the intensity of the working mode has minima, see the pump profile shown in Fig. 9(c). In this case, it is interesting to study how the maximum values of the pump amplitude and the energy of the one-mode stationary state depend on the level of additional losses $\tilde \gamma$. The numerical findings are demonstrated in Fig. 10. It is seen that the range of the pump where the lasing is monochromatic drastically increases (by an order of magnitude). So the maximum achievable energy of the working mode is much higher if the additional losses are added. Besides, a comparison of panels (a) and (b) shows that the energy of single-frequency state depends almost linearly on the pump amplitude. Also, it is important to stress that for pump amplitude $a>3$ the single-mode regime is still supported, but the dissipative terms become comparable to the conservative ones, and therefore the modes in the system are no longer clear Wannier-Stark states.

\begin{figure}[ht]
 \begin{center}
\includegraphics[width=0.35\textwidth]{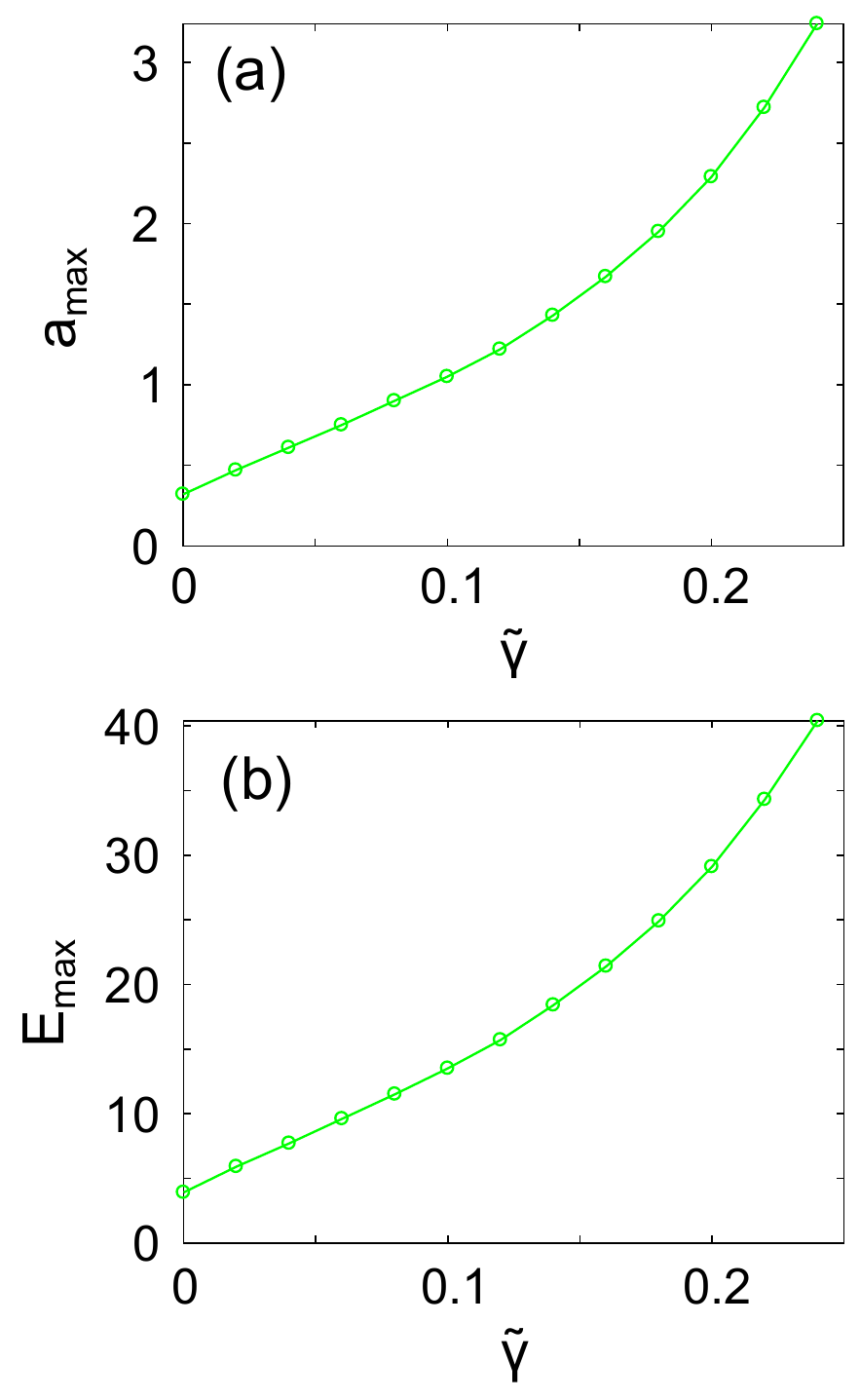}
  \caption{(Color online) (a) The dependences of the maximum pump amplitude $a_{max}$ and (b) the maximum total energy $E_{max}$ supporting a single frequency lasing regime on the value of additional losses $\tilde \gamma$ for the gain profile in Fig. 9(c). The circles correspond to the numerical data, the solid lines are guides for eyes. The nonlinear losses are nonzero only in the pumped resonators $\beta_M = \beta = 1$. Parameters are: $\mu=0.2$, $\gamma=0.01$.
  }
  \end{center}
\end{figure}

\section{Conclusion}

In this paper we considered a dynamics of Wannier-Stark (WS) states in a discrete system consisting of an array of interacting optical resonators. The presence of linear gain in some of the resonators can switch the system into lasing regime provided that the gain exceed some threshold value depending on the structure of the WS state and the spatial distribution of the pump.  
The linear gain is saturated by the nonlinear losses so that stationary states can form in the system. It is shown that in the suggested system the lasing modes are the WS states. We considered the system with the nonlinear losses present only in the pumped resonators having linear gain (only these resonators are appropriately doped) and the system with the nonlinear losses present in all resonators (all resonators are doped but the external pump profile makes linear gain to be a function of resonator index).

To study the dynamics of the system analytically a simple perturbation theory is developed. The theory allows to find the excitation thresholds, the stationary amplitudes and explains the competition between the modes. The comparison of the perturbation theory with the results of direct numerical simulation shows that the perturbation theory works well if the dissipative terms are small in the sense that they do not affect much the structure of the eigenmodes but governs the dynamics of the complex amplitudes of the modes.

It is found that the distribution of the nonlinear losses can play a crucial role in the dynamics of the excited WS states. In particular, it is shown that if linear gain exist in only one resonator then in the vicinity of the lasing threshold a single-frequency lasing regime takes place if the nonlinear losses exist in the pumped resonator only. However, if the nonlinear losses are present in all resonators then even in the vicinity of the lasing threshold the stationary state is a complex one with two dominating spectral lines corresponding to the frequencies of the two WS states having the maximum linear growth rate. This can be explain by calculating the coefficients determining the nonlinear interaction between the modes. This analysis shows that in the first case the mode of the higher amplitude successfully suppresses its competitor but in the second case this suppression is insufficient to prevent excitation of the other mode.  

The case of the linear gain present in several neighboring resonators is also considered. It is shown that single-frequency lasing is still possible in the vicinity of the lasing threshold. However, this regime occurs within a small range of pump and the energy of the lasing mode is low. If the pump amplitude grows then complex multi-frequency regime sets in. Using such a pump it is possible to reproduce Bloch oscillations (BOs) in the dissipative system. It is interesting to note that at relatively low pump amplitudes the snaking pattern manifesting BOs is symmetric. For the higher pumps the symmetry breaking occurs and the dynamics look more like two coexisting different BOs.

The problem of the stabilization of single-frequency lasing is studied. It is shown that by choosing the appropriate profile of the pump it is possible to extend the single frequency regime and to increase the energy of the lasing mode significantly. Via adding the additional losses to the system it is possible to achieve even better selection of the working mode and thus realise single frequency lasing of the chosen WS mode in a significant pump range.

The use of the WS states as the working modes can be useful, because WS modes of different frequencies have the same spatial structure just shifted in space. This fact allows to tune the frequency of the lasing mode by shifting the excitation spot. In the same time, the lasing mode is wide (includes many excited resonators), which gives an opportunity to pump it creating the gain in many resonators. So, it is possible to increase the maximum energy of the excited mode. This effect may be useful for the practical purposes, especially considering the fact, that power range of the monochromatic WS lasing can be significantly extended by the appropriate shaping of the pump.

\begin{acknowledgments}
 This work was supported by the Ministry of Science and Higher Education of Russian Federation, goszadanie no. 2019-1246.  
\end{acknowledgments}


\begin{thebibliography}{99} 

\bibitem{interest_1} Higuchi, Takuya, Mark I. Stockman, and Peter Hommelhoff. "Strong-field perspective on high-harmonic radiation from bulk solids." Physical review letters 113.21 (2014): 213901.

\bibitem{interest_2} Wang, Huan-Yu, et al. "Non-Floquet engineering in periodically driven dissipative open quantum systems." Journal of Physics: Condensed Matter 34.36 (2022): 365402.

\bibitem{interest_3} Hansen, Ursula B., et al. "Magnetic Bloch oscillations and domain wall dynamics in a near-Ising ferromagnetic chain." Nature communications 13.1 (2022): 1-8.

\bibitem{discovery_1} Wannier, Gregory H. Elements of solid state theory. CUP Archive, (1959).

\bibitem{discovery_2} Shockley, William. "Stark ladders for finite, one-dimensional models of crystals." Physical Review Letters 28.6 (1972): 349.

\bibitem{discovery_3} F. Bloch, Uber die Quantenmechanik der Elektronen in Kristallgittern, Z. Phys. 52, 555 (1929). 

\bibitem{discovery_4} C. Zener, A theory of the electrical breakdown of solid dielectrics, Proc. R. Soc. A 145, 523 (1934).

\bibitem{discovery_5}  W.V. Houston, Acceleration of electrons in a crystal lattice, Phys. Rev. 57, 184 (1940).

\bibitem{BO_exp} C. Waschke, H.G. Roskos, R. Schwedler, K. Leo, H. Kurz, and K. Kohler,  Coherent submillimeter-wave emission from Bloch oscillations in a semiconductor superlattice, Phys. Rev. Lett. 70, 3319 (1993).


\bibitem{atomic1} M.B. Dahan, E. Peik, J. Reichel, Y. Castin, and C. Salomon, Bloch Oscillations of Atoms in an Optical Potential, Phys. Rev. Lett. 76, 4508 (1996).

\bibitem{atomic2} S. Wilkinson, C. Bharucha, K. Madison, Q. Niu, and M. Raizen, Observation of Atomic Wannier-Stark Ladders in an Accelerating Optical Potential, Phys. Rev. Lett. 76, 4512 (1996).

\bibitem{atomic3} H. R. Zhang and C. P. Sun, Bloch oscillations of polaritons of an atomic ensemble in magnetic fields, Phys. Rev. A 81, 063427 (2010)

\bibitem{atomic4} Z.A. Geiger, K.M. Fujiwara, K. Singh, R. Senaratne, S.V. Rajagopal, M. Lipatov, T. Shimasaki, R. Driben, V.V. Konotop, T. Meier, and D.M. Weld, Observation and Uses of Position-Space Bloch Oscillations in an Ultracold Gas, Phys. Rev. Lett. 120, 213201 (2018).

\bibitem{atomic5} Z. Pagel, W. Zhong, R.H. Parker, C.T. Olund, N.Y. Yao, and H. Muller, Symmetric Bloch oscillations of matter waves, Phys. Rev. A 102, 053312 (2020).

\bibitem{atomic6} L. Masi, T. Petrucciani, G. Ferioli, G. Semeghini, G. Modugno, M. Inguscio, and M. Fattori, Spatial Bloch Oscillations of a Quantum Gas in a “Beat-Note” Superlattice, Phys. Rev. Lett. 127, 020601 (2021).


\bibitem{laser} S. Longhi, Dynamic localization and Bloch oscillations in the spectrum of a frequency mode-locked laser, Opt. Lett. 30, 786 (2005)

\bibitem{LC} S. Bahmani and A.N. Askarpour, Bloch oscillations and Wannier-Stark ladder in the coupled LC circuits, Phys. Lett. A 384, 126596 (2020).




\bibitem{mech1} G. Monsivais and R. Esquivel-Sirvent, Stark Ladder Resonances in Acoustic Waveguides, Journal of Mechanics of Materials and Structures, 2, 8, 1585 (2007).

\bibitem{mech2} G. Monsivais, R. Mendez-Sanchez, A. de Anda, J. Flores, L. Gutierrez, and A. Morales, Elastic Wannier–Stark Ladders in Torsional Waves
Journal of Mechanics of Materials and Structures, 2, 1629 (2007)

\bibitem{mech3} N. Lanzillotti-Kimura, A. Fainstein, B. Perrin, B. Jusserand, O. Mauguin, L. Largeau, and A. Lemaitre, Bloch Oscillations of THz Acoustic Phonons in Coupled Nanocavity Structures, Phys. Rev. Lett. 104, 197402 (2010).

\bibitem{mech4} Y.-K. Liu, H.-W. Wu, P. Hu, and Z.-Q. Sheng, Spatial Bloch oscillations in acoustic waveguide arrays, Appl. Phys. Express 14, 064501 (2021).



\bibitem{plasmonic1} A.R. Davoyan, I.V. Shadrivov, A.A. Sukhorukov, and Y.S. Kivshar, Plasmonic Bloch oscillations in chirped metal-dielectric structures, Appl. Phys. Lett. 94, 161105 (2009)

\bibitem{plasmonic2} V. Kuzmiak, S. Eyderman, and M. Vanwolleghem, Controlling surface plasmon polaritons by a static and/or time-dependent external magnetic field, Phys. Rev. B, 86, 045403 (2012)

\bibitem{plasmonic3} Bo Han Cheng, Yi-Chieh Lai, and Yung-Chiang Lan, Plasmonic Photonic Bloch Oscillations in Composite Metal–Insulator–Metal Waveguide Structure, Plasmonics, 9, 137 (2014)

\bibitem{plasmonic4} V. Kuzmiak, A. A. Maradudin, and E. R. Mendez, Surface plasmon polariton Wannier–Stark ladder, Opt. Lett. 39, 1613 (2014)

\bibitem{plasmonic5} A. Block, C. Etrich, T. Limboeck, F. Bleckmann, E. Soergel, C. Rockstuhl and S. Linden, Bloch oscillations in plasmonic waveguide arrays, Nature Communications, volume 5, Article number: 3843 (2014) 

\bibitem{plasmonic6} H. Wetter, Z. Fedorova, and S. Linden, Observation of the Wannier–Stark ladder in plasmonic waveguide arrays, Optics Letters, 47, 12, 3091 (2022)



\bibitem{exciton1} H. Flayac, D. D. Solnyshkov, and G. Malpuech, Bloch oscillations of an exciton-polariton Bose-Einstein condensate, Phys. Rev. B 83, 045412 (2011)

\bibitem{exciton2} H. Flayac, D. D. Solnyshkov, and G. Malpuech, Bloch oscillations of exciton-polaritons and photons for the generation of an alternating terahertz spin signal, Phys. Rev. B, 84, 125314 (2011)

\bibitem{exciton3} J. Beierlein, O.A. Egorov, T.H. Harder, P. Gagel, M. Emmerling, C. Schneider, S. Hofling, U. Peschel, and S. Klembt, Bloch Oscillations of Hybrid Light-Matter Particles in a Waveguide Array, Adv. Opt. Mater. 9, 2100126 (2021).


\bibitem{prediction_1} Monsivais, Guillermo, Marcelo del Castillo-Mussot, and Francisco Claro. "Stark-ladder resonances in the propagation of electromagnetic waves." Physical review letters 64.12 (1990): 1433.

\bibitem{prediction_2} De Sterke, C. Martijn, John E. Sipe, and Laura A. Weller-Brophy. "Electromagnetic Stark ladders in waveguide geometries." Optics letters 16.15 (1991): 1141-1143.

\bibitem{BO_optics1} U. Peschel, T. Pertsch,  and F. Lederer,  Optical Bloch oscillations in waveguide arrays. Opt. Lett. 23, 1701 (1998).

\bibitem{BO_optics2} A. Kavokin, G. Malpuech, A. Di Carlo, P. Lugli, and F. Rossi, Photonic Bloch oscillations in laterally confined Bragg mirrors, Phys. Rev. B 61, 4413 (2000) 

\bibitem{BO_optics3} G. Malpuech, A. Kavokin, G. Panzarini, and A. Di Carlo, Theory of photon Bloch oscillations in photonic crystals, Phys. Rev. B 63, 035108 (2001)

\bibitem{BO_optics4} S Longhi, Optical Zener-Bloch oscillations in binary waveguide arrays, Europhys. Lett. 76 416 (2006)

\bibitem{WSL_exp1} De Sterke, C. Martijn, et al. "Observation of an optical Wannier-Stark ladder." Physical Review E 57.2 (1998): 2365.

\bibitem{WSL_exp2} Ghulinyan, Mher, et al. "Zener tunneling of light waves in an optical superlattice." Physical review letters 94.12 (2005): 127401.

\bibitem{WSL_exp3} Qi, Xinyuan, et al. "Observation of accelerating Wannier–Stark beams in optically induced photonic lattices." Optics Letters 39.4 (2014): 1065-1068.

\bibitem{WSL_exp4} Mukherjee, Sebabrata, et al. "Modulation-assisted tunneling in laser-fabricated photonic Wannier–Stark ladders." New journal of physics 17.11 (2015): 115002.

\bibitem{BO_optics_exp1} T. Pertsch, P. Dannberg, W. Elflein, A. Brauer, and
F. Lederer, Optical Bloch Oscillations in Temperature
Tuned Waveguide Arrays, Phys. Rev. Lett. 83, 4752
(1999)

\bibitem{BO_optics_exp2} V. Agarwal, J. A. del Rio, G. Malpuech, M. Zamfirescu,
A. Kavokin, D. Coquillat, D. Scalbert, M. Vladimirova,
and B. Gil, Photon Bloch Oscillations in Porous Silicon
Optical Superlattices, Phys. Rev. Lett. 92, 097401 (2004)

\bibitem{BO_optics_exp3} S. Longhi, M. Lobino, M. Marangoni, R. Ramponi, P. La-
porta, E. Cianci, and V. Foglietti, Semiclassical motion
of a multiband Bloch particle in a time-dependent field:
Optical visualization, Phys. Rev. B 74 155116 (2006)

\bibitem{BO_optics_exp4} C. Bersch, G. Onishchukov, and U. Peschel, Experimen-
tal observation of spectral Bloch oscillations, Opt. Lett.
34, 2372 (2009)

\bibitem{BO_optics_exp5} Ye-Long Xu, W.S. Fegadolli, Lin Gan, Ming-Hui Lu,
Xiao-Ping Liu, Zhi-Yuan Li, A. Scherer and Yan-Feng
Chen, Experimental realization of Bloch oscillations in
a parity-time synthetic silicon photonic lattice, Nature
Communications volume 7, Article number: 11319 (2016) 

\bibitem{BO_optics_review} I.L. Garanovich, S. Longhi, A.A. Sukhorukov, and Y.S. Kivshar, Light propagation and localization in modulated photonic lattices and waveguides
Physics Reports 518 1 (2012)  


\bibitem{microlasers_review_1} Yang, Xi, et al. "Fiber optofluidic microlasers: structures, characteristics, and applications." Laser \& Photonics Reviews 16.1 (2022): 2100171.

\bibitem{microlasers_review_2} Chen, Zhi, et al. "Emerging and perspectives in microlasers based on rare-earth ions activated micro-/nanomaterials." Progress in Materials Science 121 (2021): 100814.

\bibitem{microlasers_review_3} Zhukov, A. E., et al. "Quantum-dot microlasers based on whispering gallery mode resonators." Light: Science \& Applications 10.1 (2021): 1-11.

\bibitem{microlasers_review_4} Zhang, Qing, et al. "Halide perovskite semiconductor lasers: materials, cavity design, and low threshold." Nano Letters 21.5 (2021): 1903-1914.

\bibitem{microlasers_review_5} Sumetsky, M. "Optical bottle microresonators." Progress in Quantum Electronics 64 (2019): 1-30.

\bibitem{microlasers_review_6} He, Huajun, et al. "MOF‐Based Organic Microlasers." Advanced Optical Materials 7.17 (2019): 1900077.



\bibitem{Example_model1} W. Deering, M. Molina, and G. Tsironis, Directional couplers with linear and nonlinear elements, Appl. Phys. Lett. 62, 2471, (1993).

\bibitem{Example_model2} J. Eilbeck, G. Tsironis, and S. K. Turitsyn, Stationary states in a doubly nonlinear trimer model of optical couplers, Phys. Scr. 52, 386 (1995).

\bibitem{Example_model3} P.G. Kevrekidis, K.O. Rasmussen and A.R. Bishop, The discrete nonlinear Schr\"{o}dinger Equation: a survey of recent results, International Journal of Modern Physics B 15, 2833 (2001)

\bibitem{Example_model4} N.K. Efremidis, S. Sears, D.N. Christodoulides, J.W. Fleischer, M. Segev, Discrete solitons in photorefractive optically induced photonic lattices, Phys. Rev. E 66,  046602, (2002)

\bibitem{Example_model5} A.A. Sukhorukov, Yu.S. Kivshar, H.S. Eisenberg, Y. Silberberg, Spatial optical solitons in waveguide arrays, IEEE J. Quantum Electron., 39,  31–50, (2003).
.
\bibitem{Example_model6} U. Peschel, O. Egorov, and F. Lederer, Discrete cavity solitons, Opt. Lett. 29, 1909, (2004) 

\bibitem{Example_model7} D. Pelinovsky and P. G. Kevrekidis, Stability of discrete solitons in nonlinear Schrödinger lattices, Physica D 212, 1-19, (2005).

\bibitem{Example_model8} A. Yulin, D. V. Skryabin, and P. St. J. Russell, Dissipative localized structures of light in photonic crystal films, Opt. Express, 13, 3529, (2005). 

\bibitem{Example_model9} A.V. Yulin, D.V. Skryabin , A.G. Vladimirov, Modulational instability of discrete solitons in coupled waveguides with group velocity dispersion, Opt. Express , 14, 12347, (2006).

\bibitem{Example_model10} H. Susanto, P. G. Kevrekidis, B. A. Malomed, R. Carretero-Gonzalez, and D. J. Frantzeskakis, Discrete surface solitons in two dimensions, Phys. Rev. E 75, 056605, (2007).

\bibitem{Example_model11} O.A. Egorov and F. Lederer, Lattice-cavity solitons in a degenerate optical parametric oscillator , Phys. Rev. A 76, 053816, (2007).

\bibitem{Example_model12} O.A. Egorov,  F. Lederer, and Yu.S. Kivshar, How does an inclined holding beam affect discrete modulational instability and solitons in nonlinear cavities?, Opt. Express 15, 4149 (2007).



\end{thebibliography}
\end{document}